\documentstyle[aps,prb,preprint]{revtex} 
\tighten
\begin{document}
\title{ Ground state properties of charge and magnetically frustrated  
two--dimensional quantum Josephson--junction arrays}
\date{\today}
\author{ T. K. Kope\'{c} and T. P. Polak}
\address{Institute for Low Temperature and Structure Research,
Polish Academy of Sciences,
 POB 1410, 50-950 Wroclaw 2, Poland}
 \maketitle
 \begin{abstract}
We  study a quantum Hamiltonian that models an two--dimensional
array of Josephson junctions with short range Josephson couplings,
(given by the Josephson energy $E_J$) and
charging energiey $E_C$  due to the small capacitance of the junctions.
We include the effects from  both the self-$C_0$ and the junction-$C_1$ capacitances
in the presence of external magnetic flux $f=\Phi/\Phi_0$
as well as  uniform  background of charges $q_x$.
We  derive an  effective quantum non--linear $\sigma$-model for the
array Hamiltonian which enables us a non mean--field treatment
of the zero--temperature phase transition scenario. 
We calculate the ground--state
phase  diagram, analytically deriving
$E_J^{\rm crit}(E_C,q_x,f)$ for several rational fluxes
$f=0$,$\frac{1}{2}$,$\frac{1}{3}$,$\frac{1}{4}$
and $\frac{1}{6}$
that improves upon previous theoretical treatments based on  mean--field 
approximations. 

\end{abstract} 


\section{Introduction}
  
Advances in the photolithographic micro--fabrication techniques
have experimentally allowed to manufacture Josephson--junction arrays (JJA)
with tailor specific properties. Because the junction parameters can be accurately controlled 
JJA offer a unique opportunity to test different quantum mechanical models. 
\cite{simanek}
In JJA's  the two main energy scales are set by
the Josephson coupling, $E_J$, between superconducting islands
due to Cooper pair tunneling, and to the electrostatic energy
$E_C$ (proportional to the inverse of capacitance matrix
setup by  island  self-  $C_0$ and junction- $C_1$ capacitances)
due to local deviations from charge neutrality.
When $E_J$ is much larger then $E_C$ the phases of the superconducting
order parameters on individual islands  are well defined.
 In this regime the semi-classical fluctuations determine the properties of the 
JJA. In the opposite limit, ie. for $E_C\gg E_J$, the superconducting phases  are 
dominated by strong quantum fluctuations since the Coulomb blockade pins 
the charge carriers to the islands.
Several theoretical~\cite{simani2,eckern,rojas} and experimental studies~\cite{zant}
have considered  the competition between the $E_J$ dominated phase  
and the $E_C$ dominated charging energy regions in periodic JJA.
It has been established that for 
sufficiently large charging energy the quantum phase fluctuations lead 
a complete suppression of long--range phase coherence 
even at zero temperature. This type of  {\it quantum phase transition}  has attracted 
significant interest in recent years (for a review see Ref.\onlinecite{sondhi}).

Because the junction parameters  can be controlled  accurately JJA offer a unique
system where one can  test the nature of quantum phase transitions and critical phenomena,
in particular the superconductor--insulator (SI) phase transition 
induced by quantum fluctuations.
The experimental systems can be modeled by a quantum generalization
of the classical XY model with  quantum phase of the superconducting 
order parameter on each island canonically conjugated to the excess of Cooper pair number.
The quantum parameter $\alpha=E_C/E_J$ determines the relevance of the
quantum fluctuations. Bulk of theoretical  studies
have been carried out within the self--capacitance model ($C_0\neq 0$, 
$C_1=0$), usually using different kinds of mean--field--theory or
self--consistent harmonic  approximations (for a comprehensive account
see, Ref.\onlinecite{simanek}).
 In the mutual--capacitance dominated limit
($C_1\neq 0$, $C_0=0$) the lattice model is equivalent to a
quantum Coulomb gas model,\cite{fazio} with critical properties that
are not fully understood at present. Further complication arise when {\it 
both} $C_0$ and $C_1$ are non--zero.
The charging energy can also be tuned by an applied  gate voltage $V_g$ which
control the energy required to change the number of Cooper pairs on the islands.
A uniform $V_g$ can be applied by means of gate capacitance $C_g$.\cite{roddick,sodano}
Assuming $C_g\ll C_0$ the applied voltage induces a charge $C_gV_g$ on each grain which can
be conveniently account for when one defines the a dimensionless
 charge frustration parameter $q_x=C_gV_g/2e$.
The effects of charge frustration is closely related to the so--called Bose--Hubbard
model\cite{scalettar} which describes the strongly interacting bosons under the
competing kinetic energy (related  to Josephson energy) and the potential energy
(i.e  charging energy in the JJA) effects.
The on--site (chemical) potential corresponds in this analogy to the
charge frustration parameter $q_x$ in JJA.
Frustration in quantum JJA can be also introduced by applying
magnetic field $B$. The presence of $B$ induces vortices in the system
described by the magnetic frustration parameter $f$ ($\equiv \Phi/\Phi_0$, where $\Phi$
is the magnetic flux piercing a 2D lattice plaquette and $\Phi=h/2e$
stands for the elementary flux quantum). Of special interest are
cases when $f=p/q$ with $p$ and $q$ being the rational numbers.
Particularly prominent is the case of $f=1/2$ because of the interplay
between continuous $U(1)$ symmetry of the superconducting parameter
and discrete Ising  $Z_2$ symmetry associated with antiferromagnetic
arrangements of the plaquette chiralities describing   directions of the
circulating currents in each plaquette.\cite{granato}

Usually, mean--field calculations or semiclassical approaches are not expected to be
reliable at $T=0$  and be  capable to handle spatial and quantum fluctuation effects properly
especially in two--dimensions.
Notably, there are  virtually no works beyond mean-field level on  2D JJA in the
where the physics is dominated by the
by the combination of charge $q_x$, magnetic frustration $f$ and the charging energy
related to the self-$C_0$ and junction-$C_1$ capacitances.
Recent   studies  based on perturbative expansions combine either JJA with charge
frustration and and charging energy effects\cite{choi1} or magnetic frustration and effects
of array capacitances.\cite{choi2}
Therefore, the problem we would like to address is then: What is the effect of having a competition
between the magnetic,  charge frustration and quantum effects  on the ground state
ordering in the Josephson--junction network on a square two--dimensional lattice?
The purpose of this work is therefore to study these quantum fluctuation effects in an analytical way
to understand the ground state phase diagram of the system as a function of
various control parameters like  $q_x$ and $f$ as well as the ratio of
the junction to the  self--capacitances $C_1/C_0$.
To analyze the 2D JJA beyond the mean--field theory we employ  the path--integral
formulation explicitly tailored for the macroscopic JJA Hamiltonian.
The effective action formalism allows then for an explicit implementation of the
magnetic field and offset charges effects with
the nearest--neighbor mutual and self--capacitances included.
Furthermore, we adopt an  approach based on the  quantum phase fluctuation algebra,
to map  the  microscopic JJA model Hamiltonian onto an effectively constrained system
 -- a solvable  quantum spherical model in two-dimensions that captures both quantum {\it and} 
spatial fluctuation effects beyond the mean field level.

There are four sections in this paper: Section II is 
devoted for the introduction of the microscopic Hamiltonian
of the 2D JJA  with a set of parameters  describing
with interactions and set of control parameters. 
In Sec. III the system is transformed into quantum
action  of the XY model under the magnetic field
using the ``imaginary--time" Matsubara approach.
Furthermore,  Sec. IV presents the resulting phase diagrams
and finally, Sec. V summarizes the results 
and presents some discussions. In Appendix we gave a derivation
of  closed--form analytical formulae for the  density of states of
a two-dimensional square lattice with magnetic flux
for a number of rational values $p/q$ which might be also
of general interest.

\section{Quantum phase Hamiltonian}
A Josephson junction array can be modeled by a periodic lattice
of superconducting islands separated by insulating
barriers. Each island becomes independently superconducting about the bulk
transition temperature $T_{c0}$ and it is characterized by an
order parameter $\psi({\bf r}_i)=|\psi_0({\bf r}_i)|e^{i\phi({\bf r}_i)}$, 
where ${\bf r}_i$ is a two--dimensional vector denoting the position 
of each island. The magnitude of the order parameter, $|\psi_0({\bf r}_i)|$, 
is non--fluctuating when the temperature is lowered further 
and the onset of long range phase coherence due to the tunneling 
of Cooper pairs between the  islands is responsible for the zero 
resistance drop in the arrays.
The competition  between the Josephson tunneling
and the charging (Coulomb) energy, without dissipation, 
can be modeled by the Hamiltonian
\begin{eqnarray}
{\cal H}&&={\cal H}_C+{\cal H}_J,
\nonumber\\
{\cal H}_C&&=-\frac{1}{2}\sum_{{\bf r}}
[{\bf C}^{-1}]_{{\bf r}{\bf r}'}{\hat Q}_{\bf r}{\hat Q}_{{\bf r}'},
\nonumber\\
{\cal H}_J&&=\sum_{\langle {\bf r}_1,{\bf r}_2\rangle}
J(|{\bf r}_1-{\bf r }_2|) \left[1-\cos(\phi_{{\bf r}_1}-
\phi_{{\bf r}_2})\right].
\label{hamil}
\end{eqnarray}
\narrowtext
Here, $\hat{Q}_{\bf r}=({2e}/{i}){\partial}/{\partial\phi_{\bf r}}$ is the
charge operator while ${\phi}_{\bf r}$  represents 
the superconducting phase operator of the grain at the site $\bf r$;
$J(|{\bf r}_1-{\bf r }_2|)$ is the site--dependent
Josephson coupling  and  $[{\bf C}^{-1}]_{{\bf r}{\bf r}'}$
is the inverse capacitance matrix.
The first part of the action (\ref{action}) than defines the
electrostatic energy with ${C}_{\bf  rr'}$
being a geometrical property of the array.
This matrix is normally approximated (both theoretically
and in experimental interpretation) as a diagonal (so--called
self--capacitance $C_0$) and mutual one  $C_{1}$ between nearest neighbors.
In the case of an square  2D JJA of interest here we write the capacitance
matrix as:
\begin{equation}
C_{\bf  rr'}=\left(C_0+4C_1\right)\delta_{\bf rr'}
 -C_1\sum_{\bf d}\delta_{{\bf r},{\bf r'}+ {\bf d}}
 \label{cap}
\end{equation}
with the vector ${\bf d}$ running over the nearest neighboring (n.n) islands.

\subsection{Effect of applied magnetic field}
A perpendicular magnetic field $B$ given bt the vector potential ${\bf A}$
enters the Hamiltonian (\ref{hamil}) through Peierls phase factor according to
\begin{equation}
J(|{\bf r}_i-{\bf r}_j|)\to J_B(|{\bf r}_i-{\bf r}_j|)=J(|{\bf r}_i-{\bf r}_j|)
\exp\left(\frac{2\pi i}{\Phi_0}\int_{{\bf r}_i}^{{\bf r}_j}
{\bf A}\cdot d{\bf l}  \right).
\label{phasefact}
\end{equation}
Thus, the phase shift on each junction  is determined by the vector potential ${\bf A}$
and in a typical experimental situation it can be entirely ascribed to the external field $B$.
We assume throughout this paper that the model (\ref{hamil}) is defined on a square lattice
with lattice spacing $a$. From  Eq.(\ref{phasefact}) it follows that the
properties if the array will be periodic with a period corresponding to the
one flux quantum $\Phi_0=hc/2e$ per plaquette. Of special interest are the values of the
magnetic field which corresponds to the rational values of $f=\Phi/\Phi_0$,
e.g. $f=1/2,1/4,1/4,\dots$ where $\Phi=Ba^2$. Since all properties
of the Hamiltonian (\ref{hamil}) are invariant under $f\to f+1$ and also
under $f\to-f$ it is sufficient to consider $f$ in the range $0\le f\le 1/2$.

\subsection{Offset charges}
Offset charges are an important ingredient in the experimental array
samples made of ultrasmall junctions. Several authors have shown that 
static background charges can have a pronounced effect on the SI transition
at zero temperature.\cite{Bruder,Freericks} Including the offset charges, $q_x$,
in the charging energy of Eq. (\ref{action}) gives
\begin{eqnarray}
{\cal H}_C&&\to-\frac{1}{2}\sum_{{\bf r}}
[{\bf C}^{-1}]_{{\bf r}{\bf r}'}\left({\hat Q}_{\bf r}-q_x\right)
\left({\hat Q}_{{\bf r}'}-q_x\right).
\label{hamil2}
\end{eqnarray}
Thus, offset charges, or an external gate voltage
applied between the array and the substrate, behave like a chemical
potential for injection of Cooper pairs into the
array.  
%
\section{JJA ``phase--only" action}
It is useful to derive a field-theoretic representation of 
the partition function for Eq (\ref{hamil}).
A convenient procedure is to introduce a path--integral
representation in a basis diagonal in $\phi_j$.
Given the Hamiltonian ${\cal H}$, one can simply perform a Legendre transformation to
obtain the corresponding Euclidean Lagrangian in the Matsubara ``imaginary time''
 $\tau $ formulation ($0\leq
\tau \leq 1/k_{B}T\equiv \beta $):
\begin{equation}
{\cal L}\left[ Q, \phi\right] =i\sum_{\bf r}{Q}_{\bf r}\left( \tau \right) \cdot \frac{d}{%
d\tau }\phi_{\bf r}\left( \tau \right) +{\cal H}\left[Q,\phi\right] \text{.}
\label{lagrangian}
\end{equation}
The partition function  $Z=Tre^{-{\cal H}/k_{B}T}$ of the system is then given by
\begin{eqnarray}
Z &=&\int \prod_{i}\left[ D\phi_{\bf r}\right] \int \prod_{\bf r}\left[ \frac{D%
{Q}_{\bf r}}{2\pi }\right]  e^{-\int_{0}^{\beta }d\tau {\cal L}\left[ {
Q,\phi}\right] }.  \label{PartFunction}
\end{eqnarray}
Performing the integration over the charge variables 
the partition function is expressed as
\begin{equation}
Z=\int\left[\prod_{\bf r}{\cal D}\phi_{\bf r}\right]e^{-{\cal S}[\phi]}.
\label{partfun}
\end{equation}
Here the functional integral is evaluated over the phases restricted
to the compact interval $[0,2\pi]$ and
with an effective ``phase--only" action ($\hbar=1$)
\begin{eqnarray}
{\cal S}[\phi]&=&{\cal S}_C[\phi]
+{\cal S}_J[\phi],\nonumber\\
{\cal S}_C[\phi]&=&\frac{1}{8e^2} \int_0^\beta d\tau
\sum_{\bf r,r'}
 C_{{\bf r}{\bf r}'}
\left(\frac{\partial\phi_{\bf r} }{\partial\tau}
\right)
 \left(\frac{\partial\phi_{\bf r'} }{\partial\tau}
\right)-\frac{iq_x}{2e} \int_0^\beta d\tau
\sum_{\bf r}
\left(\frac{\partial\phi_{\bf r} }{\partial\tau}
\right),
\nonumber\\
{\cal S}_J[\phi]&=&\int_0^\beta d\tau
\sum_{\langle {\bf r}_1,{\bf r}_2\rangle}
J_B(|{\bf r}_1-{\bf r }_2|) \left\{1-\cos[\phi_{{\bf r}_1}(\tau)-
\phi_{{\bf r}_2}(\tau)]\right\}.
\label{action}
\end{eqnarray}
Since the values of the phases  $\phi_i$ in Eq.(\ref{action}) which
differ by $2\pi$ are equivalent,
the path integral in Eq.(\ref{partfun}) can be written in terms of
non--compact phase variables $\theta_{\bf r}(\tau)$, 
defined on the unrestricted
interval $[-\infty,+\infty]$, and by a set
of winding numbers $\{n_{\bf r}\}=0,\pm 1,\pm 2,\dots$, which are integers
running from $-\infty$ to $+\infty$ (and physically
reflects the discreteness of the charge \cite{Bruder}),
so that $\phi_{\bf r}(\tau)=\theta_{\bf r}(0)
+2\pi  n_{\bf r}\tau/\beta +\theta_{\bf r}(\tau)$.

\subsection{Effective non--linear $\sigma$  model}
Bulk  of existing analytical works on quantum JJA have employed
variety of mean--field--like approximations
\cite{simanek,simkin,kim,zaikin} which, unfortunately, are not reliable for treatment
spatial and  temporal quantum phase fluctuations in a low dimensional systems.
To study the JJA model it appears at first natural
to use a description in terms of an effective
Ginsburg--Landau functional derived from the microscopic model
of Eq. (\ref{action}). Several studies of JJA have followed this route,
also known as the coarse grained approach first developed by Doniach.\cite{doniach}
 The essence of this method is to introduce
a complex field order parameter ${\psi}_{\bf r}$
(or equivalently a two component real field) whose expectation value
is proportional to  $\langle\exp(i\phi_{\bf r})\rangle$.
The non-zero value of this quantity
describes the ``phase-locking" or long--range phase ordering in the model.
Unfortunately, the system is governed by the Ginzburg-Landau functional
only as long as the order
parameter is small. This is a serious shortcoming of the coarse grained approach since
the method is restricted to the region of parameters in the
vicinity of the critical point. 
It is therefore desirable to extend the coarse--grained method
in a way which enables one a self--consistent description of the phase transitions
in JJA in an extended parameter range.
Following  Ref. \onlinecite{doniach2} we note  that the model (\ref{action})
encodes  phase fluctuation algebra
given by the Euclidean group $E_2$ constituted by the commutation relation
between particle $L_{\bf r}$  and phase (ladder) operators
 $P_{\bf r}=e^{i\phi_{\bf r}}$:
$[L_{\bf r},P_{\bf r'}]=-P\delta_{\bf r r'}$,
$[L_{\bf r},P^\dagger_{\bf r'}]=P^\dagger\delta_{\bf r r'}$ and $[P_{\bf r}, P_{\bf r'}]=0$
with the conserved quantity  (invariant of the $E_2$ algebra)
\begin{equation}
P_{\bf r}P^\dagger_{\bf r}=1
\label{e2constr}
\end{equation}
Thus, the proper theoretical treatment of quantum JJA
must maintain the constraint (\ref{e2constr}).
To proceed we  write the partition function
$Z=\int\left[\prod_{i}{\cal D}\phi_{i}\right]e^{-S}$ for
the model (\ref{action}) in terms  path integral representation
\cite{kleinert}
by introducing the  auxiliary complex fields ${\psi}_{i}(\tau)$
 which replace the original ladder operators $P_i$.
Furthermore, relaxing the original ``rigid" constraint (\ref{e2constr}) and imposing the
weaker  (spherical) condition $\sum_iP_{\bf r}P^\dagger_{\bf r}=N$ allows us to
formulate the problem in terms of the (exactly) soluble
quantum spherical (QS) model (see, Ref.\onlinecite{tkk}).
This can be conveniently done using the Fadeev--Popov method
with Dirac delta-functional which  facilitate both  the change
of integration variables and superimposition of the spherical constraint:
\begin{eqnarray}
Z&=&\int\left[\prod_{\bf r} {\cal D}{\psi}_{\bf r}
{\cal D}{\psi}^\star_{\bf r}\right]
\delta\left(\sum_{\bf r} |{\psi}_{\bf r}(\tau)|^2
-N\right)
e^{-{\cal S}_J[{\psi}]}\times
\nonumber\\
&\times&\int\left[\prod_{\bf r}{\cal D}\phi_{\bf r}\right]
e^{-{\cal S}_C[\phi]}
\prod_{\bf r} 
\delta\left[{\mbox Re}\,{\psi}_{\bf r}(\tau)-
{ S}_{\bf r}^x(\phi(\tau)
\right]\nonumber\\
&\times&
\delta\left[{\mbox Im }\,{\psi}_{\bf r}(\tau)-
{ S}_{\bf r}^y(\phi(\tau))
\right].
\end{eqnarray}
A convenient way to enforce the spherical constraint
is the functional analog of the $\delta-$function
representation $\delta(x)=\int_{-\infty}^{+\infty}(d\lambda/2\pi)
e^{i\lambda x}$ which introduces the Lagrange multiplier $\lambda(\tau)$
thus adding an additional quadratic term (in $\psi-$fields)
to the action (\ref{action}).
The evaluation of the effective action in terms of the $\psi$
to the second order in $\psi_{i}(\tau)$ gives
the partition function of the quantum--spherical model $Z\equiv Z_{\rm QS}$:
\begin{eqnarray}
Z_{\rm QS}
=\int\left[\prod_{\bf r} {\cal D}{\psi}_{\bf r}
{\cal D}{\psi}^\star_{\bf r}\right]
\int\left[\frac{ {\cal D}\lambda}{2\pi i}\right]
e^{-{\cal S}_{\rm QS}[
{\psi},\lambda]},
\label{statsumqsa}
\end{eqnarray}
where the action of the effective non-linear $\sigma$-model reads
\begin{eqnarray}
{\cal S}_{\rm QS}[
{\psi},\lambda]
&&= \int_0^\beta d\tau d\tau'
\left\{\sum_{\langle{\bf r}_1,{\bf r}_2\rangle}
\left[\left( {\bf J}(|{\bf r}_1-{\bf r}_2|
+\lambda(\tau)\delta_{{\bf r}_1,{\bf r}_2}\right)
\delta(\tau-\tau')\right.\right.
\nonumber\\
&&+
\left. \left.\Gamma_{02}^{-+}({\bf r}_1\tau;{\bf r}_2\tau')\right]
{\psi}^{\star}_{{\bf r}_1}(\tau)
{\psi}_{{\bf r}_2}(\tau')
-N\lambda(\tau) \delta(\tau-\tau')\right\},
\label{qsa}
\end{eqnarray}
Furthermore,
$\Gamma_{02}^{-+}({\bf r}_1\tau;{\bf r}_2\tau')
=[W_{02}^{-1}]^{-+}({\bf r}_1\tau;{\bf r}_2\tau')$
is the two--point phase vertex correlator and
\begin{equation}
W_{02}^{-+}({\bf r}_1\tau;{\bf r}_2\tau')=
\frac{1}{Z_0}\sum_{\{n_{\bf r}\}}\prod_{\bf r}\int_0^{2\pi}
 d\theta(0)\int_{\theta(0)}^{\theta(0)+2\pi n_{\bf r}}
{\cal D}\theta_{\bf r}(\tau)
e^{i[\theta_{{\bf r}_1}(\tau)-\theta_{{\bf r}_2}(\tau')]}
e^{-{\cal S}_C[\theta]},
\label{w02}
\end{equation}
with
\begin{equation}
Z_0=\sum_{\{n_{\bf r}\}}\prod_{\bf r}\int_0^{2\pi}
 d\theta(0)\int_{\theta(0)}^{\theta(0)+2\pi n_{\bf r}}
{\cal D}\theta_{\bf r}(\tau)
e^{-{\cal S}_C[\theta]},
\end{equation}
where $Z_0$ is the statistical sum of the ``non--interacting" system
described by the action ${\cal S}_C[\theta]$.
The phase--phase correlation function becomes:\cite{Bruder}
\begin{eqnarray}
W_{02}(\omega_\ell)=
\frac{8E_C}{Z_0}\sum_{q}
\frac{\displaystyle\exp\left[-4\beta E_C
(q-q_x)^2\right]}
{\displaystyle(4E_C)^2-\left[8E_C(q-q_x)	
-i\omega_\ell\right]^2},
\label{w02q}
\end{eqnarray}
where $Z_0=\sum_{q}\exp\left[-4\beta E_C
(q-q_x)^2\right]$, and the summation is performed over all
integer--valued charge states  $q=0,\pm 1,\pm 2,\dots$, which makes
the function $W_{02}(\omega_\ell)$ periodic. At low temperatures
 the sum over $q$ in Eq.(\ref{w02q})
is dominated by the charge $q$ which makes the exponent in the
numerator of  Eq.(\ref{w02q}) smallest (for $T=0$
this value is $q=0$).

In the thermodynamic limit, $N\to\infty$, we can
calculate the functional integral in Eq.(\ref{statsumqsa})
by the steepest--descent method. To proceed, we introduce propagators
associated with the order parameter field defined by
\begin{equation}
G({\bf r}_1\tau;{\bf r}_2\tau')={\langle {\psi}^{\star}_{{\bf r}_1}(\tau)
{\psi}_{{\bf r}_2}(\tau')\rangle}_{\rm QS}.
\label{opcor}
\end{equation}
The condition that the integrand in Eq.(\ref{statsumqsa})
 has a saddle point $\lambda(\tau)=\lambda_0$ is that
\begin{equation}
1=\frac{1}{N}\sum_{\bf r}
G({\bf r}\tau;{\bf r}\tau+0^+),
\label{gconstr}
\end{equation}
which becomes an implicit equation for $\lambda_0$.
The QS ensemble average is now defined by
\begin{equation}
 \langle \dots \rangle_{\rm QS} =
\frac{\int \left[\prod_{\bf r} {\cal D}{\psi}_{\bf r}
{\cal D}{\psi}^\star_{\bf r}\right] \dots
e^{  -{\cal S}_{\rm QS}[
{\psi},\lambda_0]   }}
{\int\left[\prod_{\bf r} {\cal D}{\psi}_{\bf r}
{\cal D}{\psi}^\star_{\bf r}\right]
e^{  -{\cal S}_{\rm QS}[
{\psi},\lambda_0] }}.
\end{equation}
To proceed, it is desirable to introduce  the density of states
for a 2D lattice with the magnetic flux $f=p/q$ in the form
\begin{equation}
\rho_{{p}/{q}}(E)=\frac{1}{N}\sum_{\bf k}{
\delta\left[E-\frac{J_B({\bf k})}{E_J}\right]}
\label{density}
\end{equation}
with $J_B({\bf k})$ the Fourier transform of
the Josephson interactions ${\bf J}_B(|{\bf r}_1-{\bf r}_2|)$ (cf. Eq.(\ref{phasefact}))
in a magnetic field. The problem of computing  of $\rho_{{p}/{q}}(E)$
reduces effectively to the solution of the Harper's equation\cite{harper}
relevant e.g. for tight-binding electrons on a two-dimensional lattice
with a a uniform magnetic flux per unit plaquette.\cite{hasegawa}
In  Appendix we give an analytical derivation of $\rho_{{p}/{q}}(E)$
in a closed form for several rational values of $p/q$.
A Fourier transform of Eq. (\ref{qsa})
in momentum and frequency space enables one to write
the spherical constraint (\ref{gconstr}) with the help of the density of sates
(\ref{density})  in the form
\begin{equation}
1=\frac{1}{\beta}\int_{-\infty}^{+\infty} d\xi \sum_\ell
\frac{\rho_{p/q}(\xi)}{\lambda-\xi E_J+\left[W_{02}(\omega_\ell)\right]^{-1}},
\label{2cond}
\end{equation}
where $W_{02}(\omega_\ell)$ is  the frequency transformed
phase--phase correlator of Eq. (\ref{w02})
and $\omega_\ell=2\pi\ell/\beta$ $(\ell=0,\pm 1,\pm 2,\dots)$
the (Bose) Matsubara frequencies.
As usual in a spherical model the critical behavior
and the phase transition boundary depends crucially on the
spectrum given by the density of states $\rho_{p/q}(E)$
and is determined by the denominator of the summand in the
spherical constraint equation of Eq. (\ref{gconstr}).
Specifically, when [$1/G({\bf k}=0,\omega_\ell=0)]=0$,
where $G({\bf k},\omega_\ell)$ is the Fourier transformed
order parameter correlation of Eq. (\ref{opcor}), the system displays
a critical point at
\begin{equation}
\lambda_0-\epsilon_{p/q}^{max}E_J+\left[W_{02}(\omega_\ell=0)\right]^{-1}=0,
\label{critcond}
\end{equation}
where $\epsilon_{p/q}^{max}$ is the maximum value of the  spectrum
described by the density of states (\ref{density}).
This fixes the saddle point value $\lambda$:
with  the onset of the  phase transition 
saddle point value of the Lagrange multiplier $\lambda$ ``sticks'' to that
value at criticality ($\lambda=\lambda_0$)
and stays constant in the whole low temperature phase.

\section{Phase diagrams}
By substituting the value of $\lambda_0$
from Eq.(\ref{critcond}) into (\ref{2cond}), after performing the
summation over Matsubara
frequencies, one obtains the $T\to 0$ limit the result
\begin{eqnarray}
1={\cal P}\int^{+\infty}_{-\infty}d\xi\rho_{p/q}(\xi)
\sqrt{\frac{E_C}{2(\lambda_0+2E_C-\xi E_J)}}
&&\left[{\rm sign}\left(4q_x\frac{E_C}{E_J}
+\frac{\sqrt{2(\lambda_0+2E_C-\xi E_J)E_C}}{E_J}\right)\right.
\nonumber\\
&&-\left.{\rm sign}\left(4q_x\frac{E_C}{E_J}
-\frac{\sqrt{2(\lambda_0+2E_C-\xi E_J)E_C}}{E_J}\right)\right],
\end{eqnarray}
where ${\cal P}$ denotes the principal value of the integral.
The charging energy for the two--dimensional square lattice may be explicitly
written as
\begin{eqnarray}
&&E_C=\frac{1}{2}e^2[{\bf C}^{-1}]_{\bf  r r}
\nonumber\\
&&[{\bf C}^{-1}]_{\bf r r}=
\lim_{N\to\infty}\frac{1}{N}\sum_{\bf k}\frac{1}{C_0+4C_1-2C_1(\cos 
k_x+\cos k_y)}
\nonumber\\
&&=
\frac{2}{\pi(C_0+4C_1)}{\bf K}\left(\frac{4C_1}{C_0+4C_1}  \right).
\label{total}
\end{eqnarray}
where ${\bf K}(x)$ is the elliptic integral of the first kind\cite{EllipticFunction}
(see also Eq.(\ref{ellip}) in Appendix).
Furthermore, we introduce the charging energy parameters $E_0=\frac{1}{2}e^2C_0^{-1}$ and
$E_1=\frac{1}{2}e^2C_1^{-1}$ related to the island and junction capacitances.
We fist examine the ground state phase diagram 
for a number of rational values of the flux piercing
the array without charge frustration ($q_x=0$). We give
in Fig.\ref{fig1} a graph showing the ground state phase diagram
as a function of island-to-junction charging energy for
various rational flux numbers.

The charge frustration $q_x$ may serve as a control parameter
distinguishing between ordered and non--ordered states.
The energy difference for two charge states
in each grain of the JJA with $n$ and $n+1$ extra
electrons may be reduced by changing $q_x$. As a consequence
the effects of a finite charging energy $E_C$ are weakened
and the superconducting region in the phase diagram turns out
to be enlarged. The dependence of  Josephson  to the
total charging energy Eq.(\ref{total}) $E_J/E_{C}$ is shown in Fig.\ref{fig2}
for a number of rational fluxes.
The behavior shown in in Fig.\ref{fig2} is readily understood
by realizing that the Hamiltonian (\ref{hamil}) is periodic in $q_x$
with period unity. this implies that the phase diagram should be similarly periodic,
repeating at each integer number.\cite{Bruder}

\section{Summary}

We have studied the ground phase diagram in quantum  two--dimensional 
Josephson junction arrays, using the  spherical model 
approximation with exactly evaluated density of states for
two--dimensional lattice with rational magnetic flux $f=p/q$ for a number
of values  $f=1/q$.
One of the most important issues in these systems is the role 
of the charging energy on the phase coherence transition. The detailed 
phase boundaries crucially depend on the ratio of the junction-to-self-
capacitances, $C_1/C_0$ as well as on the magnetic and charge frustration.
In arrays which are in the superconducting state at $f=0$ but
with the value $E_C/E_J$ close to the critical value, a magnetic filed
can be used to drive the array into the insulating state  (see, Fig.1).
It is interesting to compare the calculated values of $E_C/E_J$ for
$f=0$ and $f=1/2$ with the experiments:\cite{zant}
the measured critical value $E_C/E_J$ close to $T=0$ is about a factor
$0.7$ lower than the zero field value. This is to be compared with the value
$0.88$ found for $T=0$ in the present paper. Note that the critical
value of $E_C/E_J$ is greater than for others commensurate values $f=1/q$
which  suppress the superconducting state substantially. We found that
the effect of this supression is considerably larger than that
which comes  from the mean--field analysis of  JJA with magnetic flux
(for comparision see, e.g. Ref.\onlinecite{Bruder}).
Closing we note that   a reliable comparison 
with experiments presumably will still require incorporating into the model several 
other ingredients as, e.g, disorder and dissipation. These are 
topics to be considered in the future.


\appendix
\section{Density of states for 2D square lattice in a rational--flux case}

If one uses the Landau gauge, ${\bf A}=B(0,x,0)$ than the  dispersion
$\epsilon ({\bf k})$ for a square lattice of spacing $a$ with flux
$\phi=Ba^2/\Phi_0\equiv p/q$
with is given by:\cite{hasegawa}
\begin{equation}
\det \left( 
\begin{array}{cccccc}
M_{1} & -e^{ik_{x}a} & 0 & \cdots & 0 & -e^{-ik_{x}a} \\ 
-e^{-ik_{x}a} & M_{2} & -e^{ik_{x}a} & \ddots & \ddots & 0 \\ 
0 & -e^{-ik_{x}a} & M_{3} & \ddots & \ddots & \vdots \\ 
\vdots & \ddots & \ddots & \ddots & -e^{ik_{x}a} & 0 \\ 
0 & \ddots & \ddots & -e^{ik_{x}a} & M_{q-1} & -e^{ik_{x}a} \\ 
-e^{ik_{x}a} & 0 & \cdots & 0 & -e^{-ik_{x}a} & M_{q}
\end{array}
\right)=0
\label{harpereq}
\end{equation}
where 
\begin{equation}
M_{n}=-2\cos \left( k_{y}a+2\pi \phi n\right) -\epsilon\left( {\bf  k}\right).
\label{harper}
\end{equation}
Equation (\ref{harper}) is known as  Harper's equation and has been
studied extensively. If integers $p$ and $q$ are chosen to represent the flux
(with no common factor in $p$ and $q$), then the dependence on the
wave vector always appears through the
generalized structure factor  $\gamma _{n}=\cos (nk_{x}a)+\cos(nk_{y}a)$.
The density of states $\rho_\frac{p}{q}(E)$ given by the Eq.(\ref{density})
can be obtained by computing energy bands $\epsilon({\bf k})$ from
the eigenvalue equation (\ref{harpereq}). The calculation of the exact density of states
is straightforward, although for large values of $q$ may only be done
 numerically.\cite{hasegawa} However, for a number of $q$ values of interest it can be
 done analytically with a closed--form expression for $\rho_\frac{p}{q}(E)$
 as end result. Below we list these cases.
\subsection{$f=0$}

In the case of zero magnetic field the density of states for the square
two--dimensional lattice reads simply
\begin{equation}
\rho(E)=\int_{\pi/a}^{\pi/a}
\frac{d^2{\bf k}}{(2\pi)^2}\delta\left[E-\gamma_1({\bf k})\right].
\label{dens}
\end{equation}
Performing the integration over the wave-vectors we obtain
(see, Fig.\ref{d20})
\begin{equation}
\rho _{2D}\left( E\right) =\frac{1}{\pi ^{2}}{\bf  K}\left[ \sqrt{1-\left( 
\frac{E}{2}\right) ^{2}}\right] \Theta \left( 1-\frac{\left| E\right| }{2}%
\right) 
\label{2d}
\end{equation}
where
\begin{equation}
{\bf K}\left( x\right)=\int_0^{\pi/2}\frac{d\phi}{\sqrt{1-x^2\sin^2\phi}}
\label{ellip}
\end{equation}
is the elliptic integral of the first kind\cite{EllipticFunction}
and
\begin{equation}
\Theta(x)=\left\{\begin{array}{c} 1\quad {\rm for}\quad x>0\\
0\quad {\rm for}\quad x\le 0\end{array}\right.
\end{equation}
the unit step function.

\subsection{$f=\frac{1}{2}$}
For the square lattice with the half-flux quantum per
plaquette the Harper's equation (\ref{harper}) reads
\begin{equation}
\epsilon^{2}\left( {\bf  k}\right) -2\gamma _{2}-4=0
\end{equation}
so that the energy dispersion has two branches:
\begin{equation}
\begin{array}{c}
\epsilon\left( {\bf  k}\right) =\left\{ 
\begin{array}{c}
-\sqrt{4+2\left( \cos 2k_{1}+\cos 2k_{2}\right) }\text{ \ \ if \ \ }E\in
\left[ -2\sqrt{2};0\right) \\ 
+\sqrt{4+2\left( \cos 2k_{1}+\cos 2k_{2}\right) }\text{ \ \ if \ \ }E\in
\left[ 0;2\sqrt{2}\right]
\end{array}
\right.
\end{array}
\end{equation}
Integrating over the wave vectors belonging to the Brillouin zone
we obtain from Eq.(\ref{dens}) with the help of the formula (\ref{2d})
\begin{equation}
\rho _\frac{1}{2}(E)=\frac{\left| E\right| }{2}\rho _{2D}\left( \frac{E^{2}-4}{2}%
\right) {\Theta }\left( 8-E^{2}\right)
\label{rho12}
\end{equation}
Fig.\ref{d12} gives the plot of the density (\ref{rho12}).

\subsection{$f=\frac{1}{3}$}
For  $p/q=1/3$ the Harper's equation is now of cubic form
\begin{equation}
\epsilon^{3}({\bf k})-6\epsilon({\bf k})+2\gamma _{3}=0
\end{equation}
and its roots give the dispersion of the energy 
\begin{equation}
\epsilon\left( {\bf  k}\right) =\left\{ 
\begin{array}{l}
- \frac{\displaystyle1+i\sqrt{3}}{\displaystyle\left( -\gamma_3+\sqrt{-8+\gamma_3^{2}}\right) ^{\frac{1}{3}}}
   \\ +\frac{i}{2}\left( i+\sqrt{3}\right) \left( -\gamma_3+\sqrt{-8+\gamma_3^{2}}
\right) ^{\frac{1}{3}}\text{\ \ \ \ \ if \ \ }E\in \left\langle -1-\sqrt{3}
;-2\right\rangle \\ 
\frac{\displaystyle i\left( i+\sqrt{3}\right) }
{\displaystyle\left( -\gamma_3+\sqrt{-8+\gamma_3^{2}}\right)^{\frac{1}{3}}}
 \\ -   \frac{1}{2}{\displaystyle
\left( 1+i\,\sqrt{3}\right) \,
\left( -\gamma_3+\sqrt{-8+\gamma_3^{2}}\right)^{\frac{1}{3}}}
\text{\ \ \ \ if \ \ }E\in \left\langle 1-\sqrt{3};-1+
\sqrt{3}\right\rangle \\ 
\frac{\displaystyle 2+
\left( -\gamma_3+\sqrt{-8+\gamma_3^{2}}\right) ^{\frac{2}{3}}}
{\displaystyle\left( -\gamma_3+\sqrt{
-8+\gamma_3^{2}}\right) ^{\frac{1}{3}}}\text{ \ \ \ if \ \ }E\in \left\langle 2;1+
\sqrt{3}\right\rangle
\end{array}
\right.
\end{equation}
The density of states then becomes (see, Fig.{\ref{d13})
\begin{eqnarray}
\rho _\frac{1}{3}\left( E\right) &=&\frac{1}{2}\rho _{2D}\left[ \frac{%
E\left( E^{2}-6\right) }{2}\right] \left| \sqrt{\left( E^{2}-8\right) }%
\left( E^{2}-2\right) x^{\frac{1}{3}}\right| \times \\
&\times& \left\{ \left|  x^{\frac{2}{3}%
}-2\right| ^{-1}\left[ {\Theta }\left( E+1+\sqrt{3}\right) -{  %
\Theta }\left( E+2\right) \right] +  \right.
\nonumber\\
 &+&    \left|x^{\frac{2}{3}}+2\right| ^{-1}\left[
{\Theta }\left( E-1+\sqrt{3}\right) -{\Theta }\left( E\right) 
\right] + \\
&+&\left|x^{\frac{2}{3}}+2\right| ^{-1}\left[
{\Theta }\left( E\right) -{\Theta }\left( E-1+\sqrt{3}\right) 
\right] +
\nonumber\\
&+&   \left.            \left| x^{\frac{2}{3}}-2\right| ^{-1}%
\left[ {\Theta }\left( E-2\right) -{\Theta }\left( E-1-\sqrt{3}%
\right) \right] \right\},
\end{eqnarray}
where
\begin{equation}
x\left( E\right) =\frac{1}{2}\left( E^{3}-6E+\sqrt{E^{2}\left(
E^{2}-6\right) ^{2}-32}\right).
\end{equation}

\subsection{$f=\frac{1}{4}$}

The matrix formula (\ref{harper}) produces now  a four order polynomial
equation for the dispersion $\epsilon({\bf k})$.
\begin{equation}
\epsilon^{4}({\bf k})-8\epsilon^{2}({\bf k})-2\gamma _{4}+4=0
\end{equation}
Because the Eq.() is in fact bi-quadratic with respect to $\epsilon({\bf k})$
its roots can be readily found:
\begin{equation}
J\left( {\bf  k}\right) =\left\{ 
\begin{array}{c}
-\sqrt{4+\sqrt{2}\sqrt{6+\gamma _{4}}}\text{ \ \ if \ \ }E\in \left\langle -%
\sqrt{4+2\sqrt{2}};-2\sqrt{2}\right\rangle \\ 
-\sqrt{4-\sqrt{2}\sqrt{6+\gamma _{4}}}\text{ \ \ \ \ if \ \ }E\in
\left\langle -\sqrt{4-2\sqrt{2}};0\right) \\ 
+\sqrt{4-\sqrt{2}\sqrt{6+\gamma _{4}}}\text{ \ \ if \ \ }E\in \left\langle 0;%
\sqrt{4-2\sqrt{2}}\right\rangle \\ 
+\sqrt{4+\sqrt{2}\sqrt{6+\gamma _{4}}}\text{ \ \ if \ \ }E\in \left\langle 2%
\sqrt{2};\sqrt{4+2\sqrt{2}}\right\rangle
\end{array}
\right..
\end{equation}
As a result the density of states  becomes
\begin{eqnarray}
\rho _\frac{1}{4}(E) &=&\frac{\left| E^{2}-4\right| }{2}\rho _{2D}\left(\frac{%
E^{4}-8E^{2}+4}{2}\right) \times \\
&\times&\left\{ \sqrt{4+\left| E^{2}-4\right| }\left[ {\Theta }\left(
8-E^{2}\right) -{\Theta }\left( 4+2\sqrt{2}-E^{2}\right) \right] +
\right.
\nonumber\\
&+&\left.
\sqrt{4-\left| E^{2}-4\right| }{\Theta }\left( 4-2\sqrt{2}%
-E^{2}\right) \right\}.
\label{e14}
\end{eqnarray}
In Fig. \ref{d14} we have plotted the density of states given by Eq.(\ref{e14}).
\subsection{$f=\frac{1}{6}$}

The Harper' s equation  (\ref{harper}) now is given by
\begin{equation}
\epsilon^{6}({\bf k})-12\epsilon^{4}({\bf k})+24\epsilon^{2}({\bf k})-2\gamma _{6}-4=0
\label{eq16}
\end{equation}
which is a bi--cubic equation for the dispersion parameter $\epsilon({\bf k})$.
Although the solutions of  Eq.(\ref{eq16}) can be found analytically  
the result is quite involved so that
we refrain form reproducing it here. 
However, the resulting density of states (see, Fig.\ref{d16}) can be put in a closed--form expression
as follows:
\begin{eqnarray}
&&\rho _\frac{1}{6}\left( E\right)  =2^{-\frac{2}{3}}\rho _{2D}\left( \frac{%
E^{6}-12E^{4}+24E^{2}-4}{2}\right) \left| x^{\frac{1}{3}}\sqrt{y}\right| \times
\nonumber\\
&\times&\left\{ \left( \frac{2}{x}\right) ^{\frac{1}{6}}\frac{\sqrt{x^{\frac{2}{3}%
}+4\cdot \left( 2x\right) ^{\frac{1}{3}}+8\cdot 2^{\frac{2}{3}}}}{\left|
\left( 2x^{2}\right) ^{\frac{1}{3}}-16\right| }\left[ \Theta \left( 5+\sqrt{%
21}-E^{2}\right) -\Theta \left( 6+2\sqrt{3} -E^{2}\right) %
\right] \right. \nonumber \\
&&+\left| \frac{\sqrt{4\cdot 2^\frac{2}{3}-\left( 2x^\frac{1}{3}\right) ^{-1}\left[
2^{\frac{1}{3}}\left( 1-i\sqrt{3}\right) x^{\frac{2}{3}}+16\left( 1+i%
\sqrt{3}\right) \right] }}{2^{\frac{1}{3}}\left( 1-i\sqrt{3}\right) \,x^{%
\frac{2}{3}}-16\left( 1+i\sqrt{3}\right) }\right| \Theta \left( 5-\sqrt{%
21}-E^{2}\right)  \nonumber\\
&&\left. +\left| \frac{\sqrt{4\cdot 2^\frac{2}{3}-\left( 2x^\frac{1}{3}\right) ^{-1}%
\left[ 16\left( 1-i\sqrt{3}\right) +2^{\frac{1}{3}}\left( 1+i\sqrt{3}%
\right) x^{\frac{2}{3}}\right] }}{16\left( 1-i\sqrt{3}\right) -2^{\frac{1%
}{3}}\,\left( 1+i\sqrt{3}\right) x^{\frac{2}{3}}}\right|
\right.
\nonumber\\
&\times&\left.
 \left[ \Theta
(6-2\sqrt{3}-E^{2})-\Theta (2-E^{2})\right] \right\},
\end{eqnarray}
where 
\begin{eqnarray}
x(E) &=&E^{6}-12E^{4}+24E^{2}+32+\sqrt{y} \\
y(E) &=&\ \left( E^{4}-8E^{2}+8\right) ^{2}\left( E^{4}-8E^{2}-16\right)^{2}.
\end{eqnarray}
\begin{figure}
\caption{Zero temperature phase diagram for two--dimensional Josephson 
junction array on a square lattice in magnetic field  as a function
of charging energies related to self-capacitance $(E_0)$ and mutual capacitance $(E_1)$,
(normalized by the Josephson coupling energy $E_J$)
for different values of the flux ratio $f$. The inset shows closeup of the plot
for $f=0$ and $f=1/2$. The superconducting (insulting) phase
is located above (below) each curve.}
\label{fig1}
\end{figure}
\begin{figure}
\caption{Ground state phase diagram of the square JJA
as a function of the total charging energy ($E_C$,
normalized by the Josephson  coupling $E_J$) and the charge frustration
parameter $q_x$. The inset shows closeup of the plot
 for $f=0$ and $f=1/2$. The insulating phase is located  within each lobe. }
\label{fig2}
\end{figure}
\begin{figure}
\caption{Density of states  $\rho(E)$ for  two--dimensional square lattice.}
\label{d20}
\end{figure}
\begin{figure}
\caption{Density of states for  two--dimensional square lattice
with one-half of the magnetic flux quantum per plaquette. }
\label{d12}
\end{figure}
\begin{figure}
\caption{Density of states for 2D square lattice
in the magnetic field  for the flux ratio $f=1/3$.}
\label{d13}
\end{figure}
\begin{figure}
\caption{Density of states for 2D square lattice
in the magnetic field  for the flux ratio $f=1/4$.}
\label{d14}
\end{figure}
\begin{figure}
\caption{$\rho(E)$ for  2D square lattice
in the magnetic field  for the flux ratio $f=1/6$
(depicted for positive argument $E$ only, to reveal details of the band structure.}
\label{d16}
\end{figure}
%

\end{document}